\documentclass[11pt,letterpaper]{llncs}
\usepackage{authblk}
\usepackage{listings}
\usepackage{color}
\usepackage{amsmath}  
\usepackage{graphicx} 
\usepackage{amssymb} 
\usepackage{algpseudocode}
\usepackage{algorithm}
\usepackage{comment}
\usepackage{hyperref}
\usepackage{orcidlink}
\usepackage{amsfonts}
\usepackage{float}
\floatstyle{plaintop}
\restylefloat{table}
\usepackage{geometry}
\usepackage{fancyhdr}

\fancypagestyle{firstpage}{\fancyhf{}
\setcounter{page}{33}
\rfoot{\thepage}
}

\title{Anamorphic Cryptography using Baby-Step Giant-Step Recovery}
\titlerunning{Anamorphic Cryptography}

\author{
  William J. Buchanan\orcidlink{0000-0003-0809-3523}\inst{1}, Jamie Gilchrist\orcidlink{0009-0008-0649-4263}\inst{1}
}

\institute{
Blockpass ID Lab, Edinburgh Napier University, Edinburgh.
} 

\begin{document}
\maketitle
\begin{abstract} 
In 2022, Persianom, Phan and Yung outlined the creation of Anamorphic Cryptography. With this, we can create a public key to encrypt data, and then have two secret keys. These secret keys are used to decrypt the cipher into different messages. So, one secret key is given to the Dictator (who must be able to decrypt all the messages), and the other is given to Alice. Alice can then decrypt the ciphertext to a secret message that the Dictator cannot see. This paper outlines the implementation of Anamorphic Cryptography using ECC (Elliptic Curve Cryptography), such as with the secp256k1 curve. This gives considerable performance improvements over discrete logarithm-based methods with regard to security for a particular bit length. Overall, it outlines how the secret message sent to Alice is hidden within the random nonce value, which is used within the encryption process, and which is cancelled out when the Dictator decrypts the ciphertext.  It also shows that the BSGS (Baby-step Giant-step) variant significantly outperforms  unoptimised elliptic
curve methods.

\end{abstract}


\section{Introduction}
In cybersecurity, we can use anamorphic cryptography to change the viewpoint of a cipher \cite{persiano2022anamorphic}. With this, we assume that we have a dictator who will read all of our encrypted data, and will thus have a secret key of sk. The dictator (Mallory) will arrest anyone who sends secret messages that they cannot read. For this, Bob can construct a dual-decryption system, providing $sk_0$ to the Dictator and $sk_1$ to Alice (or Alice can supply Bob with one she chooses as $sk_1$ herself). As far as Mallory knows, he has the only key for the ciphertext, and outwardly the encrypted data appears standard (i.e has no difference in structure to that of a normal encrypted message). This paper outlines a new method of creating anamorphic cryptography using Baby-Step Giant-Step Recovery to provide a fast recovery method for Alice, and which improves on existing approaches. 

\section{Related work}

An implementation of the ElGamal method for anamorphic cryptography is given in \cite{banfi2024anamorphic}. While discrete logarithm methods have been used to implement anamorphic cryptography \cite{githubRobustanamorphicencryptionelgamalpyMain}, they tend to be slow in their operation.  Dodis et al. \cite{dodis2025anamorphic} liken the approach of anamorphic cryptography to the addition of backdoors into semantically secure schemes, where entities might be forced to hand over their decryption keys. This technique offers the ability to send covert data or to later claim plausible deniability for actions. Carnemolla et al. \cite{carnemolla2025anamorphic} have outlined that there are certain schemes which are \emph{anamorphic resistant} and that, in some cases, anamorphic encryption is similar to substitution attacks. The methods that support anamorphic encryption include RSA-OAEP, Pailler, Goldwasser-Micali, ElGamal schemes, Cramer-Shoup, and Smooth Projective Hash-based systems. 

Chu et al. \cite{chu2025threshold} investigated cases where the dictator is involved in key generation, and where it is still possible to implement anamorphic communication. This involves the usage of threshold signature schemes, and the adversary is included within the signing group.

\section{Method}
This paper outlines the integration of ElGamal methods  \cite{elgamal1985public} with ECC for the implementation of anamorphic encryption. With anamorphic encryption, we can have a public key of $pk$ and two private keys of $sk0$ and $sk1$. Bob can then have two messages of:
\begin{align}
m_0=\textrm{"I love the Dictator"}\\
m_1=\textrm{"I hate the Dictator"}
\end{align}

Bob then encrypts the two messages with the public key( the $PK_0$ of the Dictator):

\begin{align}
CT=Enc(pk,m_0, msg_1)
\end{align}

The Dictator will then decrypt with $sk_0$ and reveal the first message:

\begin{align}
Dec(sk_0,CT) \rightarrow m_0
\end{align}

Alice will decrypt with her key and reveal the second message:
\begin{align}
Dec(sk_1,CT) \rightarrow m_1
\end{align}

And, so, the Dictator thinks that they can decrypt the message, and gets, “I love the Dictator”. Alice, though, is able to decrypt the ciphertext to a different message of “I hate the Dictator”.

\subsection{ElGamal encryption}
With ElGamal encryption using elliptic curves \cite{asecuritysite_82778}, Alice generates a private key ($x$) and a public key of:

\begin{align}
Y=x.G
\end{align}

 and where $G$ is the base point on the curve. She can share this public key ($Y$) with Bob. When Bob wants to encrypt something for Alice, he generates a random value ($r$) and the message value ($M$) and then computes:

\begin{align} 
C_1 = r.G\\
C_2 = r.Y+M
\end{align}

To decrypt, Alice takes her private key ($x$) and computes:

\begin{align} 
M=C_2-x.C_1
\end{align} 

This works because:
\begin{align} 
M=C_2-y.C_1 = r.x.G+M - x.r.G=M
\end{align} 

\begin{figure*}
    \centering
    \includegraphics[width=1.0\textwidth]{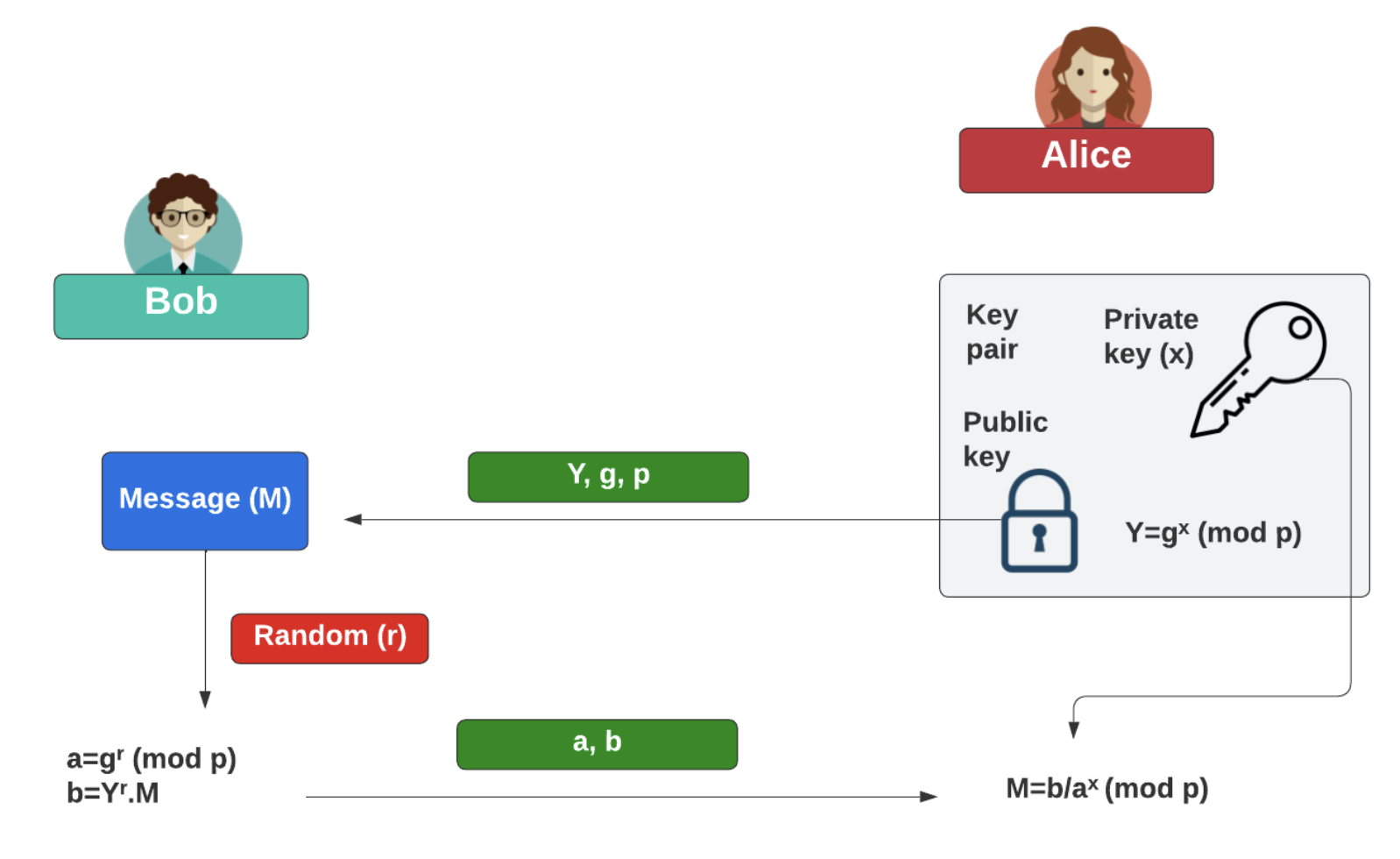}
    \caption{ElGamal encryption with ECC}
    \label{fig:ecc}
\end{figure*}

Figure \ref{fig:ecc} outlines how Bob can encrypt data for Alice.

\subsection{Anamorphic Encryption with elliptic curves}
First, we pick a curve, such as secp256k1, which has a base point of $G$. Bob can then pick a secret key for the Dictator of $sk_{Dictator}$. The public key is then:
\begin{align} 
pk = sk_{Dictator}.G
\end{align} 

Bob then generates a random scalar value of $t$ and takes the secret message of $cm$, and produces:
\begin{align} 
r=cm+t
\end{align} 

The value of $t$ will be Alice's secret key. To encrypt the message of $m$, Bob uses:
\begin{align} 
rY = r.pk\\
rG = r.G\\
rYval =\textrm{Int}(rY)\\
c0 := rYval+M\\
c1 := rG\\
\end{align} 

The cipher is the ($c_0,c_1$). To decrypt by the Dictator:

\begin{align} 
yC = sk_{Dictator}.c1\\
yC_{val} = \textrm{Int}(yC)\\
res_{Dictator} = c_0- yC_{val}\\
\end{align} 

Alice can then decrypt with her key ($t$):

\begin{align} 
tc= t.G\\
res_{Alice} = c_1-tc
\end{align} 

Alice will then search through the possible values of $res_{Alice}$ to find the value of $cm$ that matches the elliptic curve point. This works because:

\begin{align}
res_{Alice} = c_1-tc = r.G - t.G = r.G - (r-cm).G = cm.G 
\end{align}

It is worth mentioning that both the $t$ and $tc$ for Alice must be treated as secret. This is because anyone who has access to the Public key and to the $c1$ component of the ciphertext can do:

Unlike traditional Public Key Cryptogtaphy, in the case of Alice, both the private key $t$ and the Public key($t_c$) must be treated as secrets. This is in contrast to normal PKI, where a public key can be shared safely. The reason for this is that anyone with Alices Public Key, can then also brute force the cm. They can do : 

\begin{align}
cm \cdot G = c_1 - tc
\end{align}

Given $cm \cdot G$, recovering $cm$ reduces to solving a discrete logarithm problem:
\begin{align}
cm = \log_G (cm \cdot G)
\end{align}
However, although the ECDLP is considered computationally difficult for large ranges, the system intentionally restricts $cm$ to a low-bit space (e.g., 30 bits) so it can be recovered in a reasonable time on lower-performance hardware. This small search space then makes it feasible for anyone with $t_c$ and $c_1$ to brute-force the remaining 30 bits of entropy. 
Some sample code and a test run are given in the appendix.

\subsection{Baby-Step Giant-Step}

Anamorphic cryptography offers a compelling approach to covert communication and plausible deniability, but it does come with specific security considerations. 

In traditional public key cryptography, the public key can be treated as such, public. However, in this scheme, Alice's public and private keys ($t$ and $tc$ must be treated as secrets. These can effectively be seen as 'dual keys', where each must be safeguarded from disclosure to other parties. This is because an attacker who intercepts the ciphertext $(c_0, c_1)$ and learns $t \cdot G$ ($tc$) can compute:

\begin{align}
cm = \log_G(c_1 - t_c)
\end{align}

This works because cm is small (30 bits for example), and thus Alice is able to brute-force or Baby-Step Giant-Step $cm$ in a reasonable time on commodity hardware. 

For anyone else who does not know $t$ or $t_c$, computing $cm$ from $c_1$ is equivalent to solving the elliptic curve discrete logarithm problem(ECDLP). If $t_c$ is secret, and the size of $cm$ is small (even just two bits), without $t$ or $t_c$, the attacker is left with only $c_1 = r  \cdot G$, which is completely infeasible within modern computing.

In this setup, both $t$ and $tc = t \cdot G$ must be kept private. If $tc$ leaks, anyone with access to $c_1$ can compute $c_1 - tc = cm \cdot G$. Since $cm$ is intentionally small (e.g., 30 bits), this makes brute-force recovery viable. In effect, $tc$ acts like a second private key — and its exposure would compromise the covert message. Thus, in the conventional sense the keys( either $t$ or $t_c$) Alice holds are both private keys. 

\subsection{Implementation}
Now we will outline the implementation of our solution. Algorithm 1 outlines the method of applying anamorphic using ECC encryption, and Algorithm 2 defines the method for the Dictator to decrypt. Algorithm 3 provides the method for Alice to decrypt the message that is passed to her. With this, Alice will reveal the covert message, and which cannot be seen by the Dictator.

\begin{algorithm}
\caption{Anamorphic ECC Encryption}
\begin{algorithmic}[1]
\Require Public key $pk = sk_0 \cdot G$, cover message $m_0$, covert message $cm \in \mathbb{Z}$, Alice's secret key $t \in \mathbb{Z}$
\Ensure Ciphertext $(c_0, c_1)$
\State $r \gets cm + t$
\State $r_G \gets r \cdot G$
\State $r_Y \gets r \cdot pk$
\State $r_Y^{\text{val}} \gets$ Integer encoding of $r_Y$
\State $c_0 \gets r_Y^{\text{val}} + m_0$
\State $c_1 \gets r_G$
\State \Return $(c_0, c_1)$
\end{algorithmic}
\end{algorithm}

\begin{algorithm}
\caption{Decryption by Dictator}
\begin{algorithmic}[1]
\Require Secret key $sk_0$, ciphertext $(c_0, c_1)$
\Ensure Decrypted message $m_0$
\State $y_C \gets sk_0 \cdot c_1$
\State $y_C^{\text{val}} \gets$ Integer encoding of $y_C$
\State $m_0 \gets c_0 - y_C^{\text{val}}$
\State \Return $m_0$
\end{algorithmic}
\end{algorithm}

\begin{algorithm}
\caption{Covert Decryption by Alice}
\begin{algorithmic}[1]
\Require Alice's secret key $t$, ciphertext component $c_1$, generator $G$
\Ensure Covert message $cm$
\State $tc \gets t \cdot G$
\State $res \gets c_1 - tc$ \Comment{Now $res = cm \cdot G$}
\State Search (e.g., brute-force or BSGS) for $cm$ such that $cm \cdot G = res$
\State \Return $cm$
\end{algorithmic}
\end{algorithm}

\subsection{Key Derivation and Use as a Shared Secret}

An alternative application can involve treating the result of Alice's decryption as a shared elliptic curve point:

\begin{align}
res_{Alice} = c_1 - t \cdot G = cm \cdot G
\end{align}

This point can act as a symmetric secret index or be hashed to derive encryption keys, enabling integration with secure storage systems or encrypted messaging layers.

\subsection{Secrecy of Alice’s Public Key}

It is critical that Alice’s scalar $t$ and corresponding $t \cdot G$ remain confidential. If an adversary gains access to the latter, they can compute:

\begin{align}
cm \cdot G = c_1 - t \cdot G \Rightarrow cm = \log_G (cm \cdot G)
\end{align}

Given that $cm$ is intentionally small (e.g., 30-bit), this discrete log is trivially solvable, compromising the covert message. Therefore, $t \cdot G$ must be treated as secret in this scheme, contrary to standard public key conventions.

\section{Evaluation}
In order to benchmark the performance and practical efficiency of the anamorphic system, we have run a series of tests on a Windows PC with an Intel i7 3770, 3.4~GHz with 16GB of DDR3 RAM, where we compare both the DLP and ECDLP variants, while also showing optimisations. We then solve for $cm$ of varying sizes up to 34 bits. For the unoptimised version of both the DLP and ECDLP variants, tests were done up to 20 bits only due to the time taken. For optimised versions, we tested up to the full 34 bits in both cases. 

Each script used in testing is included in the Appendix (Section~\ref{sec:appendix}).

\begin{itemize}
    \item \textbf{Vanilla-DLP}: A non-optimised discrete logarithm (DLP) based implementation using naive scalar multiplication.
    \item \textbf{ECC-DLP-Vanilla}: A basic elliptic curve variant using a similar brute-force search for the scalar.
    \item \textbf{BSGS-DLP}: An optimised DLP implementation leveraging the Baby-Step Giant-Step (BSGS) algorithm.
    \item \textbf{ECCDLP-BSGS}: An optimised elliptic curve implementation using BSGS with curve precomputations.
\end{itemize}

\begin{table}
\centering
\caption{Alice Decryption Time vs. Message Index ($cm$) for Different Schemes}
\label{tab:alice_decryption}
\begin{tabular}{p {4cm}p{4cm} p{4.5cm}}
\hline
\textbf{Scheme} & \textbf{cm} & \textbf{Decryption Time (ms)} \\
\hline\hline
Vanilla-DLP & 9 & 6.9701 \\
Vanilla-DLP & 99 & 7.5207 \\
Vanilla-DLP & 999 & 6.1685 \\
Vanilla-DLP & 9,999 & 63.6082 \\
Vanilla-DLP & 99,999 & 1700.6238 \\
Vanilla-DLP & 999,999 & 27769.2745 \\
\hline
ECC-DLP-Vanilla & 9 & 6.5441 \\
ECC-DLP-Vanilla & 99 & 42.0362 \\
ECC-DLP-Vanilla & 999 & 429.0826 \\
ECC-DLP-Vanilla & 9,999 & 4,280.2626 \\
ECC-DLP-Vanilla & 99,999 & 42,637.5404 \\
ECC-DLP-Vanilla & 999,999 & 442,947.4792 \\
\hline
BSGS-DLP & 9 & 4.6448 \\
BSGS-DLP & 99 & 6.9718 \\
BSGS-DLP & 999 & 8.4206 \\
BSGS-DLP & 9,999 & 7.9787 \\
BSGS-DLP & 99,999 & 12.5426 \\
BSGS-DLP & 999,999 & 14.5983 \\
BSGS-DLP & 9,999,999 & 12.0785 \\
BSGS-DLP & 99,999,999 & 6.8217 \\
BSGS-DLP & 999,999,999 & 11.2848 \\
BSGS-DLP & 9,999,999,999 & 10.1387 \\
\hline
ECCDLP-BSGS & 9 & 1.5771 \\
ECCDLP-BSGS & 99 & 4.5974 \\
ECCDLP-BSGS & 999 & 10.1518 \\
ECCDLP-BSGS & 9,999 & 47.3780 \\
ECCDLP-BSGS & 99,999 & 145.2986 \\
ECCDLP-BSGS & 999,999 & 458.9628 \\
ECCDLP-BSGS & 9,999,999 & 1,487.1940 \\
ECCDLP-BSGS & 99,999,999 & 4,693.6531 \\
ECCDLP-BSGS & 999,999,999 & 14,142.6721 \\
ECCDLP-BSGS & 9,999,999,999 & 43893.8668 \\
\hline
\end{tabular}
\end{table}

The ECCDLP-BSGS variant significantly outperforms its unoptimised elliptic curve counterpart (ECC-DLP-Vanilla), reducing decryption time by over three orders of magnitude for large $cm$ values. The DLP variants also benefit markedly from BSGS optimisation, with decryption times reduced from tens of seconds to milliseconds. Interestingly, the optimised \textbf{BSGS-DLP} implementation outperforms the \textbf{ECCDLP-BSGS} variant in raw decryption time across most $cm$ values, owing to faster integer arithmetic compared to elliptic curve point operations.

\section{Conclusions}
While RSA-OAEP, Pailler, Goldwasser-Micali, ElGamal schemes, Cramer-Shoup, and Smooth Projective Hash-based systems all support anamorphic cryptography, the usage of elliptic curve methods provides an opportunity to enhance the overall performance of the methods implemented for the ElGamal technique. The results show that the   ECCDLP-BSGS variant significantly outperforms  unoptimised elliptic
curve methods for anamorphic cryptography.

\bibliographystyle{IEEEtran}
\bibliography{main}

\section{Appendix}\label{sec:appendix}

This appendix outlines the main algorithms used within the anamorphic ECC encryption system described in the paper. Full source code and command line tools that support this implementation are available at: \\
\texttt{\url{https://github.com/billbuchanan/babydictator}}.

\subsection{Source Code and Tools}

The full implementation is provided in the companion GitHub repository. Tools include:
\begin{itemize}
  \item \texttt{keygen.go} – Generates key pairs for both the Dictator and Alice
  \item \texttt{encrypt.go} – Encrypts both cover and covert messages
  \item \texttt{decrypt\_dictator.go} – Decrypts the cover message using $sk_0$
  \item \texttt{decrypt\_alice.go} – Recovers the covert message using Alice’s key
\end{itemize}

The repository also includes benchmarking scripts and a chatbot interface to demonstrate how covert communication works in real-time using this proof of concept (POC).

\subsection{Application Implementation}

To support further experimentation, testing and real-time demo, we have implemented a full-stack application for Anamorphic Encryption. This consists of Go CLI tools, a Python API wrapper, and a Chatbot front-end. Collectively, these represent a messenger service, where covert data can be hidden in the ciphertext.

\subsection{Command-Line Tools (CLI)}

The core cryptographic logic is implemented in Go and exposed through standalone CLI binaries. These tools form the computational backend and include:

\begin{itemize}
    \item \texttt{keygen.go} — Generates Dictator and Alice key pairs
    \item \texttt{encrypt.go} — Performs anamorphic encryption with standard and covert message support
    \item \texttt{decrypt\_dictator.go} — Decrypts messages with the Dictator's key
    \item \texttt{decrypt\_alice.go} — Recovers covert message $cm$ using Alice's key and optional BSGS optimisation
\end{itemize}

Each tool supports flexible inputs (via flags), and all test cases used for benchmarking (see Section~\ref{tab:alice_decryption}), other than the DLP-based tests, were produced using these scripts.

\subsection{REST API Layer}
A Python Flask-based web API wraps the Go CLI binaries, exposing endpoints for key generation, encryption, and both decryption flows. This API layer supports:

\begin{itemize}
    \item Stateless operation with ephemeral or uploaded keys
    \item Endpoint support for JSON-based requests:
        \begin{itemize}
            \item \texttt{POST /api/keygen}
            \item \texttt{POST /api/encrypt}
            \item \texttt{POST /api/decrypt-alice}
            \item \texttt{POST /api/decrypt-dictator}
        \end{itemize}
    \item Docker-friendly deployment and integration with the frontend
\end{itemize}

\subsection{Repository Access and Scripts}

The full source code, CLI tools, Python API server, React frontend, schema definitions are hosted at:
https://github.com/test-sum/ecc-anamorphic-api-covert-chat


Scripts include:
\begin{itemize}
    \item CLI: \texttt{keygen.go}, \texttt{encrypt.go}, \texttt{decrypt\_alice.go}, \texttt{decrypt\_dictator.go}
    \item API: \texttt{app.py} wrapping CLI binaries with RESTful endpoints
    \item Frontend: Chatbot UI for encryption testing and message interpretation
    \item Examples: Full usage walkthroughs, pre-built binaries, and LLM integration hooks
\end{itemize}

\section{Authors}
\noindent {\bf William (Bill) J Buchanan OBE FRSE }  is a Professor of Applied Cryptography in the School of Computing, Edinburgh and the Built Environment at Edinburgh Napier University. He is a Fellow of the BCS and a Principal Fellow of the HEA. Bill was appointed an Officer of the Order of the British Empire (OBE) in the 2017 Birthday Honours for services to cybersecurity, and,  in 2024, he was appointed as a Fellow of the Royal Society of Edinburgh (FRSE). In 2023, he received the "Most Innovative Teacher of the Year" award at the Times Higher Education Awards 2023 (the "Oscars of Higher Education"), and was awarded “Cyber Evangelist of the Year” at the Scottish Cyber Awards in 2016 and 2025. Along with this, he has won the Best Lecturer/Tutor for Computing at the Student-voted Excellence Awards six times. Bill lives and works in Edinburgh and is a believer in fairness, justice, and freedom.  His social media tagline reflects his strong belief in changing the world for the better: "A Serial Innovator. An Old World Breaker. A New World”. \\

\noindent{\bf Jamie Gilchrist} is a self-taught cryptography and cybersecurity enthusiast with over 20 years of professional experience in the IT industry. His research interests include applied cryptography, decentralised systems, identity solutions and the intersection of Machine Learning, privacy and data security.

\end{document}